\documentstyle[prl,aps,epsf,multicol]{revtex}
\begin{document}

\title{Phase transition of the one-dimensional coagulation-production process}
\author{G\'eza \'Odor}
\address{Research Institute for Technical Physics and Materials Science, \\
H-1525 Budapest, P.O.Box 49, Hungary}    
\maketitle

\begin{abstract}
Recently an exact solution has been found \cite{HH} for the $1d$ 
coagulation production process: $2A\to A$, $A\text{\O}A\to 3A$ with 
equal diffusion and coagulation rates. This model evolves into the 
inactive phase independently of the production rate with $t^{-1/2}$
density decay law. 
Here I show that cluster mean-field approximations and Monte Carlo 
simulations predict a continuous phase transition for higher 
diffusion/coagulation rates as considered in \cite{HH}.
Numerical evidence is given that the phase transition universality
agrees with that of the annihilation-fission model with low diffusions.
\end{abstract}

\begin{multicols}{2}

One-dimensional, non-equilibrium phase transitions have been 
found to belong to a few universality classes, the most robust 
of them is the directed percolation (DP) class 
\cite{Dick-Mar,Hin2000}. 
According to the hypothesis of \cite{Jan81,Gras82} all continuous phase 
transitions to a single absorbing state in homogeneous 
systems with short ranged interactions belong to this class
provided there is no additional symmetry and quenched 
randomness present. 

Recent studies on the annihilation fission (AF or PCPD) process 
$2A\to\text{\O}$, $2A\to 3A$ \cite{HT97,Carlon99,Hayepcpd,Odo00}
found evidence that there is a phase transition in this model
that does not belong to any known universality classes.
This model without the diffusion of single particles
was introduced originally by \cite{IJen93l}.
The renormalization group analysis of the corresponding 
bosonic field theory was given by \cite{HT97}.
This study predicted a non-DP class transition, 
but it could not tell to which universality class this 
transition really belongs.
An explanation  based on symmetry arguments are still missing
but numerical simulations suggest \cite{HayeDP-ARW,Odo00}
that the behavior of this system can be well described 
(at least for strong diffusion) by coupled sub-systems: 
single particles performing annihilating random walk 
coupled to pairs ($B$) following DP process:
$B\to 2B$, $B\to\text{\O}$. 
The system has two non-symmetric
absorbing states: one is completely empty, in the other 
a single particle walks randomly. 
Owing to this fluctuating absorbing state this model does
not oppose the conditions of the DP hypothesis.
Some exponents are close to those of the PC class 
\cite{Dick-Mar,Hin2000} but the order parameter exponent 
($\beta$)  has been found to be very far away 
from both of the DP and PC class values \cite{Odo00}. 
In fact this system does not exhibit neither a $Z_2$ 
symmetry nor a parity conservation that appear in models
with PC class transition.
It is conjectured  \cite{HH} that this kind of phase transition 
appears in models where (i) solitary particles diffuse,
(ii) particle creation requires two particles and (iii)
particle removal requires at least two particles to meet.

In this paper the following one-dimensional coagulation production 
processes will be investigated:

a) Spatially symmetric coagulation production processes:
\begin{eqnarray}
A\text{\O}A\stackrel{f}{\longrightarrow}3A , \\
2A\stackrel{c/2}{\longrightarrow}A\text{\O},  \ \ \  
2A\stackrel{c/2}{\longrightarrow}\text{\O}A , \\
A\text{\O}\stackrel{d}{\leftrightarrow}\text{\O}A 
\end{eqnarray}

b) Spatially asymmetric coagulation production processes:
\begin{eqnarray}
AA\text{\O}\stackrel{f/2}{\longrightarrow}3A, \ \ \
\text{\O}AA\stackrel{f/2}{\longrightarrow}3A ,  \\
2A\stackrel{c/2}{\longrightarrow}A\text{\O},  \ \ \  
2A\stackrel{c/2}{\longrightarrow}\text{\O}A , \\
A\text{\O}\stackrel{d}{\leftrightarrow}\text{\O}A
\end{eqnarray}
Both versions fulfil conditions (i-iii.) but Henkel et al. \cite{HH} 
show that for $d=c$ the symmetric version always evolve into the 
inactive state with $\rho\propto t^{-0.5}$ scaling law.
They argue that the asymmetric version displays a non-equilibrium
phase transition. The difference is said to be similar to the hard-core
effects observed in one-dimensional models 
\cite{Dhar,dimercikk,arw2cikk,Park,barw2cikk}.
Hard-core particle exclusion effects can really change both the dynamic
\cite{Dhar,dimercikk,arw2cikk} and static \cite{Park,barw2cikk} 
behavior of one dimensional systems by introducing blockades into the
particle dynamics but in this work I argue that not this kind of 
hard-core effects responsible for the lack of phase transition.

One can quickly check by simulations that for $d \le c$ the density
in the asymmetric version decays in much the same way 
-- with $\rho\propto t^{-0.5}$ scaling law -- as in case of the symmetric 
version. Furthermore I shall show that if the coagulation rate is smaller
than the diffusion rate particles can escape before removal an active phase 
will emerge with a continuous phase transition belonging to same class 
that was found in the AF model for weak diffusion.
Therefore both version exhibit qualitatively the same phase diagram. 

To prove this first I shall apply cluster mean-field approximations 
(GMF) \cite{gut87,dic88}, which can predict phase diagrams qualitatively 
well. The mean-field equation for the steady state of both version is
\begin{equation}
0 = f (1-p_A) p_A^2 - c p_A^2 \label{MFe}
\end{equation}
where $p_A$ is the probability of $A$-s at a given site. 
Note that the diffusion rate $d$ does not play a role in this
approximation. By introducing the parametrization $c=p(1-d)$,
$f=(1-p)(1-d)$ -- that is similar to that of the PCPD model --
this has the solution:
\begin{equation}
\rho = p_A = \frac{2p-1}{p-1} \label{MFsol}
\end{equation}
for $p<1/2$ and $\rho=0$ if $p\ge 1/2$. Therefore an active state
appears in the mean-field approximation already.

\begin{figure}[h]
\epsfxsize=70mm
\epsffile{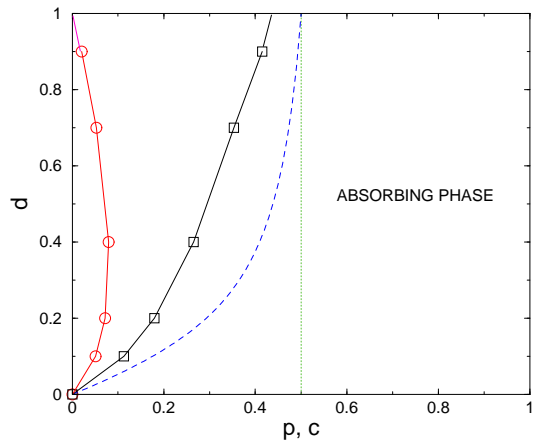}
\vspace{2mm}
\caption{Phase diagram of the symmetric coagulation-production model.
Dotted line : mean-field approximation, dashed line : pair approximation,
squares: simulation results. The circles show $d_c$ as the function of $c$.
Lines connecting symbols are used to guide eyes only.
}
\label{phased}
\end{figure}

For higher order cluster mean-field approximations 
similar scenario can be found, but one has to treat the two versions
separately. The density in pair approximation for the symmetric 
version is:
\begin{equation}
\frac{\left(p -1\right) \,p^2 - 2\,d\,p\,\left(p^2+2\,p-2\right)  + 
d^2\,\left(p^3+5\,p^2+4\,p-4\right) }
{\left(p-1\right) \,p^2-2\,d\,p\,\left(p^2+2\,p-2\right)  + 
d^2\,\left(p^3+5\,p^2-4\right) }
\end{equation}
One can easily prove that if the coagulation rate is equal to the 
diffusion rate $d=p(1-d)/2$ this gives a single $\rho=0$ absorbing 
state solution in agreement with \cite{HH}. 
The steady state solution with positive density is possible if 
\begin{equation}
d  > \frac{p^2-p}{p^2+3p-2}  \  \ .
\end{equation}
This gives the phase boundary in pair approximation that is a 
continuous unlike in case of the PCPD model \cite{Carlon99}
(see Fig.\ref{phased}).
The pair density in this approximation 
\begin{equation} 
c = \frac{{\left( \left(p -1\right) \,p - 
       d\,\left(p^2 + 3\,p -2 \right)  \right) }^2 \left( p-1\right)^{-1}}
       { \left( p-1\right) \,p^2 - 2\,d\,p\,\left( p^2 + 2\,p -2 \right)  + 
      d^2\left( p^3 + 5\,p^2 -4 \right) }
\end{equation}
has a leading order singularity all along the phase transition line 
\begin{equation}
c \propto (p_c-p)^2
\end{equation}
suggesting one universality class unlike in the case of PCPD model
\cite{Carlon99}.

\begin{figure}[h]
\epsfxsize=70mm
\epsffile{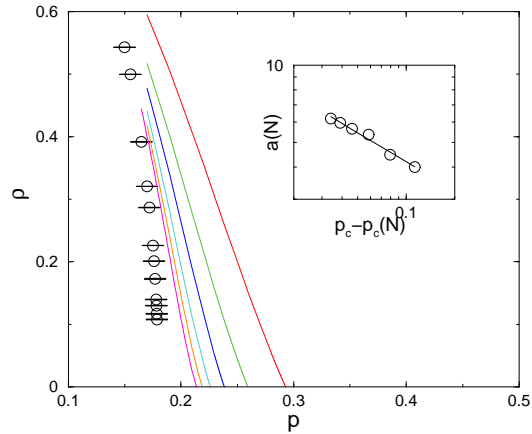}
\vspace{2mm}
\caption{Cluster mean-field approximations in the symmetric 
coagulation-production model for $d=0.2$.
The curves correspond to steady state density solutions as the
function of $p$, for $N=2,3,4,5,6,7$ (right to left). 
The circles with error bars represent simulation results.
Insect: Corresponding coherent anomaly amplitudes with a
power-law fitting.
}
\label{rho-an}
\end{figure}

The GMF solutions for $N=3,4,5,6,7$ block sizes have been determined 
numerically at $d=0.2$. The approximation level is constarined by the
numerical stability of the fixed point solution in the 
multi-dimensional space of $N$-block probability variables.
As Figure \ref{rho-an} shows the $\rho_N$ density curves of 
different approximations converge to the simulation results.

Using these data an estimate can be given for the order parameter 
density exponent $\rho\propto |p-p_c|^{\beta}$ using the Coherent anomaly 
method (CAM) \cite{suz86}, which has been proven to give precise 
estimates for the DP \cite{GMFDP} and PC \cite{MeOd95} classes. 
According to CAM the amplitudes $a(N)$ of the cluster mean-field 
singularities scale in such a way that
\begin{equation}
a(N) \propto |p_c(N)-p_c|^{\beta - \beta_{MF}} \label{anoscal}
\end{equation}
the exponent of true singular behavior can be estimated.
From the mean-field solution (\ref{MFsol}) one read-off that 
$\beta_{MF}=1$. The critical point $p_c$ can be estimated either
by extrapolating on the GMF results or by simulations.
Linear extrapolation at $d=0.2$ for $p_c(1/N\to 0)$ gives: 
$p_c=0.182(2)$. Monte Carlo simulations on large systems 
-- discussed below -- give a more precise estimate: 
$p_c=0.17975(8)$. The amplitudes $a(N)$ near $p_c(N)$ are 
determined by linear fitting from the $\rho_N(p)$ data
and shown in insect of Fig.\ref{rho-an}. as the function 
of $p_c(N)-p_c$.
A power-law with exponent $\beta - \beta_{MF}=-0.43(3)$ can fairly 
well applied for points corresponding to $N>2$ approximations
giving an estimate: $\beta=0.57(3)$, which agrees well with former
results for the AF model with small diffusion rates
\cite{Odo00}.

\begin{figure}[h]
\epsfxsize=70mm
\epsffile{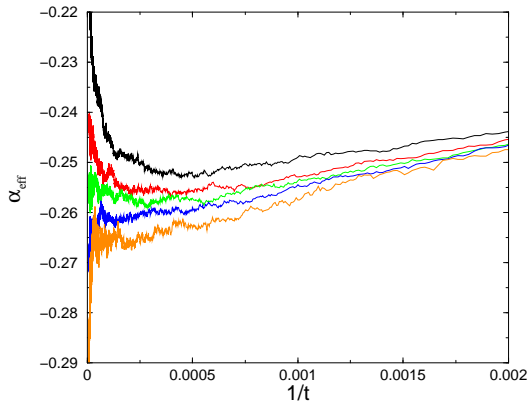}
\vspace{2mm}
\caption{Local slopes of the density decay in the symmetric coagulation production
process. Different curves correspond to 
$p=0.1795$, $0.1797$, $0.1798$, $0.1799$ $0.18$ (from bottom to top).
Throughout the whole paper $t$ is measured in units of 
Monte-Carlo sweeps (MCS).}
\label{dec}
\end{figure}

Monte-Carlo simulations of the symmetric process started 
from fully occupied lattices of size $L=40000$ 
show a phase transition for $d=0.2$ and $p_c=0.17975(10)$ 
(see Fig.\ref{dec}).
The local slopes of the density decay:
\begin{equation}
\alpha_{eff}(t) = {- \ln \left[ \rho(t) / \rho(t/m) \right] 
\over \ln(m)} \label{slopes}
\end{equation}
(where we use $m=8$ usually) at the critical point go to 
exponent $\alpha$ by a straight line, while in 
sub(super)-critical cases they veer down(up) respectively. 
For the critical point ($p_c=0.17975(8)$) one can estimate that
the effective exponent tends to $\alpha = 0.263(9)$, which agrees 
with results for the AF model \cite{Hayepcpd,Odo00} again.
For other $d$-s similar results have been found.

In the supercritical region the steady states have been determined for 
different $\epsilon=p-p_c$ values. Following level-off the densities 
were averaged over $10^4$ MCS and $1000$ samples.
By looking at the effective exponent defined as
\begin{equation}
\beta_{eff}(\epsilon_i) = \frac {\ln \rho(\epsilon_i) -
\ln \rho(\epsilon_{i-1})} {\ln \epsilon_i - \ln \epsilon_{i-1}} \ \ ,
\end{equation}
\begin{figure}[h]
\epsfxsize=70mm
\epsffile{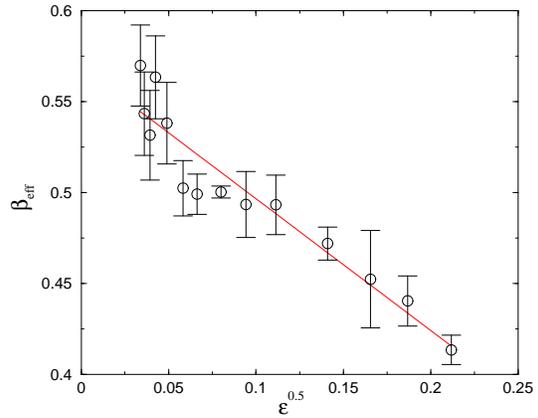}
\vspace{4mm}
\caption{Effective order parameter exponent results. Linear extrapolation 
results in $\beta=0.57(1)$.}
\label{beta}
\end{figure}
one can read-off: $\beta_{eff}\to\beta \simeq 0.57(1)$, which is in
good agreement with the exponent of the AF model for weak diffusion
determined by coherent anomaly method and simulations \cite{Odo00}.

The simulations and the cluster mean-field approximations show that if 
the diffusion rate is lowered this phase transition disappears and the 
system will decay with the $\rho\propto t^{-0.5}$ law independently 
of $f$ in {\bf both} versions. As expected the asymmetric version 
exhibits a phase transition with the same universal properties as the
symmetric version. Example for $d=0.2$ the transition point is at 
$c=0.359(1)$, $f/2=0.4409(1)$ with the decay exponent $\alpha=0.27(1)$.

In conclusion coagulation production models exhibit a phase transition 
if the diffusion is fast enough.
The spatial symmetry of the production process has been found to be
irrelevant as in case of the AF process \cite{Odo00}. 
The critical behavior agrees well with that of the AF model in its
weak diffusion rate region. An open question is that why can one not see
the cyclically coupled behavior in this model similarly as in the
PCPD model as $d\to1$.
The corrections to scaling are getting very strong in this limit
that make numerical solutions very confusing, but one has to realize 
that the $B\to\text{\O}$ process of pairs (present in AF) 
is missing in this model. Therefore a {\it single} universality class 
in this model and {\it two distinct classes} in the AF model is likely.
This conjecture is strengthened by the pair mean-field results :
one obtains analytically the {\it same} singular behavior here 
and {\it two distinct singular behavior} in case of the PCPD model 
along the phase transition line.

\begin{table}
\caption{Summary of results}
\label{tab}
\begin{tabular}{|l|r|r|r|r|}
$d$          &  0.1         & 0.2        & 0.4      &  0.7\\
\hline
$p_c$        & 0.1129(1)    & 0.17975(8) & 0.2647(1)& 0.3528(2)\\ 
$\alpha$     &  -           & 0.263(9)   & 0.268(8) & 0.275(8)\\
$\beta$      &  -           & 0.57(1)    & 0.58(1)  & 0.57(1)\\
$\beta_{CAM}$&  -           & 0.57(3)    &   -      &     -     \\
\end{tabular}
\end{table}

\vspace{3mm}
\noindent
{\bf Acknowledgements:}\\

The author would like to thank N\'ora Menyh\'ard for stimulating
discussions and Malte Henkel, Haye Hinrichsen for their comments.
Support from Hungarian research 
fund OTKA (Nos. T-25286 and T-23552) and from B\'olyai 
(No. BO/00142/99) is acknowledged.

\end{multicols}
\end{document}